\documentstyle[preprint,aps]{revtex}
\begin{document}
\title{Decoupled oscillations and resonances of three neutrinos in matter}
\author{G. Cheng\thanks{%
Email: gcheng@ustc.edu.cn}}
\address{CCAST (World Laboratory), P.O. Box 8730, Beijing 100080, P.R. China%
\\
and\\
Astronomy, and Applied Physics Department, USTC., \\
Hefei, Anhui, 230026, P.R. China\thanks{%
Mailing address}}
\maketitle

\begin{abstract}
\baselineskip20pt In a previous paper, we gave a new theoretical framework
in which three neutrino mixing in matter are discussed. Rigorous analytical
solutions are obtained. In present paper an approximate method is developed
for studying three flavor neutrino oscillations in matter in the new
framework. Using condition of $\Delta m_{1}^{2}\ll \Delta m_{2}^{2}$,
decoupled resonances, which is discovered by Kuo {\it et al.} previously, are
obtained without small angle approximation. Calculation and approximation is
consistent and rigorous. All the formulae appear in simple symmetric form
and some physical characteristic are more explicit. Around the two resonant
points, the mass eigenvalue whose eigenfunction does not take part in the
oscillation process, can be easily calculated to higher precision. We find,
the oscillation amplitude is dominated by vacuum mixing and is smaller than
1 for smaller A resonance, but it is not in the larger A. This is
characteristic for the three flavor oscillations, but do not exist in two
flavor.
\end{abstract}



\newpage \baselineskip20pt

\section{Introduction}

Recently new experimental results for neutrino appear rapidly \cite{1}. More
and more theoretical articles are published attempting at fixing their mass
and mixing angles \cite{2}. Solar neutrino and some other experiments relate
to the mixing and oscillations of the neutrino in matter \cite{1,2,3}. For
studying these data, theoretical methods for solving the propagation and
oscillation equations of neutrino in matter are required. Many important
results have been obtained for two flavors. An important fact, discovered
more than ten years ago, is that there is resonance of the mixing angle in
matter due to the interaction of charged current \cite{4}. Because of the
increase of mixing parameters the situation is more complex in three flavors
than two. An important advance was gained in the research of MSW oscillation
in three flavors in 1986. It was found that the three flavor oscillations
are approximately decoupled into two independent oscillations and resonances
between two flavors \cite{5}. Many important results from the researches of
two flavors oscillation can be used in the three flavor case \cite{6}. Many
works have been done and papers are published. They have been used in
analyzing the experimental results, analytically and numerically \cite{6,7}.
However a new theoretical framework and analytical method may be helpful to
this task. We have made a series of efforts in this respect. In a previous
paper, the author gave a new theoretical framework to discuss three neutrino
mixing in matter \cite{8}. Results are given analytically in explicit,
symmetric, and very simple form. Now we apply them to the approximate
solutions and discuss three flavor oscillations in matter.

In section II, the results obtained in the first paper \cite{8} and will be
used in the paper, are quoted.

In section III, using and only using the condition of $\Delta m_{1}^{2}\ll
\Delta m_{2}^{2},$ a strict approximate mathematical method is developed to
deal with the eigenvalue equation of mass in matter for three neutrinos. By
the method, a cubic algebra equation is reduced into two of quadratic, in
two regions around $A\sim \Delta m_{1}^{2}$ and $A\sim \Delta m_{2}^{2}$
respectively. Thus, we get four solutions at each region, but only three of
them satisfies our condition imposed on eigenvalues used for reducing the
equation. In this way, decoupled oscillations and resonances can be read out
directly. In addition, the solution which does not take part in the
resonance can be calculated to higher precision which will be very useful in
derivating propagating equation in a followed paper.

In section IV, a consistent rigorous mathematical derivation is presented
using our theoretical framework \cite{8} to expose the behaviors of the
mixing parameters in matter around $A\sim \Delta m_{1}^{2}$ and $A\sim
\Delta m_{2}^{2}$ when $\Delta m_{1}^{2}\ll \Delta m_{2}^{2}.$ Approximately
decoupled oscillation and resonance behavior are obtained between $U
_{e1}^{m}$ and $U _{e2}^{m}$ around the $A\sim \Delta m_{1}^{2},$ and
between $U _{e2}^{m}$ and $U _{e3}^{m}$ around $A\sim \Delta m_{2}^{2}$
respectively. This phenomena is the partner of the oscillation and resonance
between cos$\theta ^{m}$ and sin$\theta ^{m}$ in two flavors \cite{4}. In three
flavor, this had been discovered as the decoupled resonances of $\theta
_{1}^{m}$ around $\Delta m_{1}^{2}$ and $\theta _{3}^{m}$ around $\Delta
m_{2}^{2}$ by T.K. Kuo and his collaborator previously \cite{5}. However,
there are some phenomena which are not discussed or proved explicitly
before. In the lower resonance, the amplitude, $\sqrt{U _{e1}^{m2}+U
_{e2}^{m2}}\approx \sqrt{\eta _{1}^{2}+\eta _{2}^{2}}<1.$ It is determined by
the vacuum parameters. However, in the higher resonance, $\sqrt{U
_{e2}^{m2}+U _{e3}^{m2}}\approx 1$ which is independent on the vacuum
mixing parameters. There is notable difference between two flavor and three
flavor oscillation and resonance in matter, even if in the case of decoupled
resonances in three flavors. When the vacuum mixing angles are not small,
the new observation is meaningful.

In the end, in section V, the results are discussed and the conclusion is
given.

\ 

\section{Theoretical framework}

\ We begin from the general equation of the eigenvalve problem and some of
the rigorous analytical solutions obtained in a previous paper \cite{8}. All
symbols and definitions used in that paper will be used in here without
explanation.. The equation dominated the eigenvalues of three neutrinos in
matter is. 
\begin{equation}
\lambda ^{3}+a\lambda ^{2}+b\lambda +c=0
\end{equation}%
where 
\begin{equation}
a=-(A+\Delta m_{2}^{2})
\end{equation}%
\begin{equation}
b=-(\Delta m_{1}^{2})^{2}+\left[ \Delta m_{1}^{2}\left( \eta _{11}-\eta
_{22}\right) +\Delta m_{2}^{2}\left( \eta _{11}+\eta _{22}\right) \right] A
\end{equation}%
and 
\begin{equation}
c=(\Delta m_{1}^{2})^{2}\Delta m_{2}^{2}-\Delta m_{1}^{2}\left[ \Delta
m_{2}^{2}\left( \eta _{11}-\eta _{22}\right) -\Delta m_{1}^{2}\eta _{33}%
\right] A
\end{equation}%
The solutions for $\nu _{e}$\ mixing matrix elements in matter are 
\begin{equation}
U_{eu}^{m}=N^{u}\sum\limits_{i=1}^{3}\frac{\eta _{i}^{2}}{\lambda
_{u}-\Delta _{i}}\qquad u=1,2,3
\end{equation}%
where $N^{u}$ is the normalizing constant.%
\begin{equation}
N^{u}=\left[ \sum\limits_{i=1}^{3}\left( \frac{\eta _{i}}{\lambda
_{u}-\Delta _{i}}\right) ^{2}\right] ^{-\frac{1}{2}}\qquad u=1,2,3
\end{equation}%
and $\Delta _{i}s$ are 
\begin{equation}
\begin{array}{l}
\Delta _{1}=-\frac{1}{2}\left( m_{2}^{2}-m_{1}^{2}\right) =-\Delta m_{1}^{2}
\\ 
\Delta _{2}=\frac{1}{2}\left( m_{2}^{2}-m_{1}^{2}\right) =\Delta m_{1}^{2}
\\ 
\Delta _{3}=m_{3}^{2}-\frac{1}{2}\left( m_{2}^{2}+m_{1}^{2}\right) =\Delta
m_{2}^{2}%
\end{array}%
\end{equation}%
The mass eigenvalues in matter expressed by $\lambda _{u}$ are 
\begin{equation}
M_{u}^{2}=\lambda _{u}+\frac{1}{2}\left( m_{1}^{2}+m_{2}^{2}\right) 
\end{equation}%
Remembering that we have proved 
\begin{equation}
\eta _{1}^{2}+\eta _{2}^{2}+\eta _{3}^{2}=1
\end{equation}%
it is easy to see 
\begin{equation}
U_{e1}^{m2}+U_{e2}^{m2}+U_{e3}^{m2}=1
\end{equation}%
This relation is helpful to our task below.\ Now we make use of these
formulae and the condition $\Delta m_{1}\ll \Delta m_{2}$ to derive
approximate solutions in which decoupled resonances emerged explicitly.%
\vspace{10pt}

\section{Approximate solutions of eigenvalues}

\subsection{Dimensionless Equation\protect\vspace{10pt}}

Before we begin work, it is convenient to make the equation dimensionless in
favor of estimating the quantitative order for each term in the
Eq.(1)-Eq.(4). Taking $\lambda ,\,\Delta m_{1}^{2}$ and $\Delta m_{2}^{2}$
in the unit of $A$, we introduce the following variable\ transformation. 
\begin{equation}
\overline{\lambda }=\frac{\lambda }{A},\quad \overline{\Delta m}_{1}^{2}=%
\frac{\Delta m_{1}^{2}}{A}\quad and\quad \overline{\Delta m}_{2}^{2}=\frac{%
\Delta m_{2}^{2}}{A}
\end{equation}%
We get 
\begin{equation}
\overline{\lambda }^{3}+\overline{a}\overline{\lambda }^{2}+\overline{b}%
\overline{\lambda }+\overline{c}=0
\end{equation}%
where 
\begin{equation}
\overline{a}=-(1+\overline{\Delta m}_{2}^{2})
\end{equation}%
\begin{equation}
\overline{b}=-(\overline{\Delta m}_{1}^{2})^{2}+\left[ \overline{\Delta m}%
_{1}^{2}\left( \eta _{11}-\eta _{22}\right) +\overline{\Delta m}%
_{2}^{2}\left( \eta _{11}+\eta _{22}\right) \right] 
\end{equation}%
and 
\begin{equation}
\overline{c}=(\overline{\Delta m}_{1}^{2})^{2}\overline{\Delta m}_{2}^{2}-%
\overline{\Delta m}_{1}^{2}\left[ \overline{\Delta m}_{2}^{2}\left( \eta
_{11}-\eta _{22}\right) -\overline{\Delta m}_{1}^{2}\eta _{33}\right] 
\end{equation}

Three flavor neutrino oscillations can be decoupled approximately into two
independent oscillations between two flavors and resonances, if\ the
neutrino masses have the hierarchical characteristic, or more exactly if $%
\Delta m_{1}^{2}\ll \Delta m_{2}^{2},$ where $\Delta m_{1}^{2}=\left(
m_{2}^{2}-m_{1}^{2}\right) /2$ and $\Delta m_{2}^{2}=m_{3}^{2}-\left(
m_{2}^{2}+m_{1}^{2}\right) /2.$ This fact have been observed by T.K.Kuo and
his collaborator previously. Now, we derive it in this new theoretical
framework.\vspace{10pt}

\subsection{Solutions around $A\sim \Delta m_{1}^{2}$\protect\vspace{10pt}}

First, we search for the approximate solutions around the region $A\sim
\Delta m_{1}^{2}$. In present case, 
\begin{equation}
\overline{\Delta m}_{1}^{2}\sim 1\quad and\quad \overline{\Delta m}%
_{2}^{2}\gg 1
\end{equation}%
\vspace{10pt}

\subsubsection{Solution to the small eigenvalues $\protect\lambda \sim
\Delta m_{1}^{2}.$\protect\vspace{10pt}}

We try to search for the solutions which satisfy $\lambda \sim \Delta m_{1}^{2}$%
. Then $\overline{\lambda }\sim \overline{\Delta m}_{1}^{2}\sim 1.$
Neglecting all the terms which is of order 1 in Eq.(12)-Eq.(15), we obtain
the following approximate equation, 
\begin{equation}
\overline{\lambda }^{2}-\left( \eta _{11}+\eta _{22}\right) \overline{%
\lambda }-\left[ (\overline{\Delta m}_{1}^{2})^{2}-\overline{\Delta m}%
_{1}^{2}\left( \eta _{11}-\eta _{22}\right) \right] =0
\end{equation}%
There are two solutions for it 
\begin{equation}
\overline{\lambda }_{\mp }=\frac{\left( \eta _{11}+\eta _{22}\right) \mp 
\sqrt{\left( \eta _{11}+\eta _{22}\right) ^{2}+4\left[ (\overline{\Delta m}%
_{1}^{2})^{2}-\overline{\Delta m}_{1}^{2}\left( \eta _{11}-\eta _{22}\right) %
\right] }}{2}
\end{equation}%
Both of them satisfy the condition{\bf \ }$\overline{\lambda }\sim \overline{%
\Delta m}_{1}^{2}$ $\sim 1$. Thus we obtain two solutions 
\begin{equation}
\lambda _{1,2}=\frac{A\left( \eta _{11}+\eta _{22}\right) \mp \sqrt{%
A^{2}\left( \eta _{11}+\eta _{22}\right) ^{2}+4\left[ (\Delta
m_{1}^{2})^{2}-A\Delta m_{1}^{2}\left( \eta _{11}-\eta _{22}\right) \right] }%
}{2}
\end{equation}%
\vspace{10pt}

\subsubsection{Solution to the large eigenvalue $\protect\lambda \sim \Delta
m_{2}^{2}.$ \protect\vspace{10pt}}

We try to search for the third solution as $\lambda \sim \Delta m_{2}^{2}.$
That is the solution satisfying $\overline{\lambda }\sim \overline{\Delta m}%
_{2}^{2}\gg 1.$ We can neglect the terms whose orders are equal or smaller
than $\overline{\Delta m}_{2}^{2}$ in Eq.(12)-Eq.(15). We get an approximate
equation as 
\begin{equation}
\overline{\lambda }^{2}-\left( 1+\overline{\Delta m}_{2}^{2}\right) 
\overline{\lambda }+\overline{\Delta m}_{2}^{2}\left( \eta _{11}+\eta
_{22}\right) =0
\end{equation}%
There are two solutions 
\begin{equation}
\overline{\lambda }_{\pm }=\frac{1}{2}\left[ 1+\overline{\Delta m}%
_{2}^{2}\pm \sqrt{\left( 1+\overline{\Delta m}_{2}^{2}\right) ^{2}-4%
\overline{\Delta m}_{2}^{2}\left( \eta _{11}+\eta _{22}\right) }\right] 
\end{equation}%
but, only the solution $\overline{\lambda }_{+}$ satisfies $\overline{%
\lambda }\sim \overline{\Delta m}_{2}^{2}$ which is used for obtaining the
reduced equation. Therefore we get a maximum solution 
\begin{equation}
\lambda _{3}=\frac{1}{2}\left[ A+\Delta m_{2}^{2}+\left( A+\Delta
m_{2}^{2}\right) \sqrt{1-\frac{4A\Delta m_{2}^{2}\left( \eta _{11}+\eta
_{22}\right) }{\left( A+\Delta m_{2}^{2}\right) ^{2}}}\right] 
\end{equation}%
Remaining to the order $A,$ we get 
\begin{equation}
\lambda _{3}\approx \Delta m_{2}^{2}+A\eta _{33}
\end{equation}%
{\bf \vspace{10pt}}

\subsection{Solutions around $A\sim \Delta m_{2}^{2}.$\protect\vspace{10pt}}

When $A\sim \Delta m_{2}^{2}$ 
\begin{equation}
\overline{\Delta m}_{2}^{2}\sim 1\quad and\quad \overline{\Delta m}%
_{1}^{2}\ll 1
\end{equation}%
\vspace{10pt}

\subsubsection{Larger solution $\protect\lambda \sim \Delta m_{2}^{2}.$%
\protect\vspace{10pt}}

We try to search for the solutions it satisfy $\overline{\lambda }\sim 
\overline{\Delta m}_{2}^{2}.$ Neglecting all the terms having the order
smaller than $\left( \overline{\Delta m}_{2}^{2}\right) ^{2}$, from
Eq.(12)-Eq.(15), we obtain an approximate equation 
\begin{equation}
\overline{\lambda }^{2}-(1+\overline{\Delta m}_{2}^{2})\overline{\lambda }+%
\overline{\Delta m}_{2}^{2}\left( \eta _{11}+\eta _{22}\right) \lambda =0
\end{equation}%
There are two solutions for it 
\begin{equation}
\overline{\lambda }_{\mp }=\frac{1}{2}(1+\overline{\Delta m}_{2}^{2})\mp 
\frac{1}{2}\sqrt{(1+\overline{\Delta m}_{2}^{2})^{2}-4\overline{\Delta m}%
_{2}^{2}\left( \eta _{11}+\eta _{22}\right) }
\end{equation}%
Both of them satisfy the condition $\lambda \sim \Delta m_{2}^{2}$. Thus we
obtain 
\begin{equation}
\lambda _{2,3}=\frac{1}{2}(A+\Delta m_{2}^{2})\mp \frac{1}{2}\sqrt{(A+\Delta
m_{2}^{2})^{2}-4A\Delta m_{2}^{2}\left( \eta _{11}+\eta _{22}\right) }
\end{equation}%
\vspace{10pt}

\subsubsection{Smaller solution $\protect\lambda \sim \Delta m_{1}^{2}$%
\protect\vspace{10pt}}

We try to search for the third solution in{\bf \ }$\overline{\lambda }\sim 
\overline{\Delta m}_{1}^{2}\ll 1.$ In Eq.(12)-Eq.(15), neglecting the terms
whose order are equal or smaller $\left( \overline{\Delta m}_{1}^{2}\right)
^{2},$ we obtain an approximate equation 
\begin{equation}
\left( \eta _{11}+\eta _{22}\right) \overline{\lambda }-\overline{\Delta m}%
_{1}^{2}\left( \eta _{11}-\eta _{22}\right) =0
\end{equation}%
Its solution is 
\begin{equation}
\overline{\lambda }=\overline{\Delta m}_{1}^{2}\frac{\eta _{11}-\eta _{22}}{%
\eta _{11}+\eta _{22}}
\end{equation}%
That is 
\begin{equation}
\lambda _{1}=\Delta m_{1}^{2}\frac{\eta _{11}-\eta _{22}}{\eta _{11}+\eta
_{22}}
\end{equation}%
In fact, in our present method, higher precision can be reached for this
solution by neglecting only the terms in order $\left( \overline{\Delta m}%
_{1}^{2}\right) ^{3}.$ Now we have a twice algebra equation 
\begin{equation}
\begin{array}{ll}
(1+\overline{\Delta m}_{2}^{2})\overline{\lambda }^{2}-\left[ \overline{%
\Delta m}_{1}^{2}\left( \eta _{11}-\eta _{22}\right) +\overline{\Delta m}%
_{2}^{2}\left( \eta _{11}+\eta _{22}\right) \right] \overline{\lambda }- & 
\\ 
\qquad \text{\quad }-\left[ (\overline{\Delta m}_{1}^{2})^{2}-\overline{%
\Delta m}_{1}^{2}\left( \eta _{11}-\eta _{22}\right) \right] \overline{%
\Delta m}_{2}^{2}+\left( \overline{\Delta m}_{1}^{2}\right) ^{2}\eta _{33} & 
=0%
\end{array}%
\end{equation}%
There are two solutions for this equation, but only one of them satisfies $%
\overline{\lambda }\sim \overline{\Delta m}_{1}^{2}$ which is used to reduce
the equation. It is the smaller one. The solution is a minimum solution for $%
A\sim \Delta m_{2}^{2}$. When we discuss propagating equation in a followed
paper, this higher precision solution is very useful, but we are not
necessary write it here explicitly.{\bf \vspace{10pt}}

\section{Approximately decoupled resonance behaviors of $U_{\lowercase{e}1}^{m}$, $U%
_{\lowercase{e}2}^{m}$, $U_{\lowercase{e}3}^{m}$}

In section II, we have introduced a set of solutions, Eq.(5) for the $%
U_{e,u}^{m}$ from a previous paper \cite{8}, because we are interested only
in the mixing associated with $\nu _{e}$ in matter. They are expressed as
functions of the vacuum parameters and potential $A.$ Now we use the
equation to discuss the approximately decoupled resonant behaviors of $%
U_{e1}^{m},$ $U_{e2}^{m},$ and $U_{e3}^{m}.${\bf \vspace{10pt}}

\subsection{\noindent Resonant form for $\protect\lambda _{1}$\ and $\protect%
\lambda _{2}$\ around $A\sim \Delta m_{1}^{2}$.\protect\vspace{10pt}}

When $A\sim \Delta m_{1}^{2},$ we have get three approximate solutions. They
are expressed in Eq.(19) and Eq.(22). There is a possible resonance between
eigenstates of $\lambda _{1}$ and $\lambda _{2}$ when the quantity in the
square root in Eq.(19) takes the minimum value. It can be reached when the
following condition of $A=A_{l}$ is satisfied: 
\begin{equation}
A_{l}=2\Delta m_{1}^{2}\frac{\eta _{11}-\eta _{22}}{\left( \eta _{11}+\eta
_{22}\right) ^{2}}
\end{equation}%
Introducing 
\begin{equation}
\rho _{l}=A\left( \eta _{11}+\eta _{22}\right) -2\Delta m_{1}^{2}\frac{\eta
_{11}-\eta _{22}}{\eta _{11}+\eta _{22}}\quad and\quad \delta _{l}=\Delta
m_{1}^{2}\frac{2\eta _{1}\eta _{2}}{\eta _{11}+\eta _{22}}
\end{equation}%
we can rewrite the solutions Eq.(19) as 
\begin{equation}
\lambda _{1,2}=\frac{1}{2}\left[ A\left( \eta _{11}+\eta _{22}\right) \mp 
\sqrt{\rho _{l}^{2}+4\delta _{l}^{2}}\right] \quad A\sim \Delta m_{1}^{2}
\end{equation}%
It is clear that $4\delta _{l}^{2}>0$ when $\eta _{1}\eta _{2}\neq 0.$ There
is a possible resonance between eigenstates of $\lambda _{1}$ and $\lambda
_{2}$ when the condition $A=A_{l}$ is satisfied. Because $\lambda _{3}\sim
\Delta m_{2}^{2}\gg \lambda _{1,2}\sim \Delta m_{1}^{2},$ this resonance is
decoupled approximately from $\lambda _{3}.$

Taking the traditional representation and let $U=U_{2}U_{3}U_{1}$ \cite{8}, we have 
\begin{equation}
\frac{\eta _{11}-\eta _{22}}{\left( \eta _{11}+\eta _{22}\right) ^{2}}=\frac{%
c_{1}^{2}c_{3}^{2}-s_{1}^{2}c_{3}^{2}}{\left(
c_{1}^{2}c_{3}^{2}+s_{1}^{2}c_{3}^{2}\right) ^{2}}=\frac{\cos 2\theta _{1}}{%
c_{3}^{2}}
\end{equation}%
Thus, the lower resonance condition, Eq.$\left( 32\right) ,$ becomes 
\begin{equation}
A=A_{l}=2\Delta m_{1}^{2}\frac{\cos 2\theta _{1}}{c_{3}^{2}}
\end{equation}%
It is the same as the result obtained by Kuo {\it et al.}\vspace{10pt}

\subsection{The behaviors of $U_{e1}^{m}$, $U_{e2}^{m}$\ \ and $U_{e3}^{m}$\
in the neighborhood of $A\sim \Delta m_{1}^{2}.$\protect\vspace{10pt}}

We can use the equations of $U_{e1}^{m}$ and $U_{e2}^{m}$ and take the
approximation as follows 
\begin{equation}
U_{e1,2}^{m}\approx N^{1,2}\left( \frac{\eta _{1}^{2}}{\lambda _{1,2}+\Delta
m_{1}^{2}}+\frac{\eta _{2}^{2}}{\lambda _{1,2}-\Delta m_{1}^{2}}\right)
\end{equation}%
Correspondingly, we have 
\begin{equation}
N^{1,2}\approx \left[ \left( \frac{\eta _{1}}{\lambda _{1,2}+\Delta m_{1}^{2}%
}\right) ^{2}+\left( \frac{\eta _{2}}{\lambda _{1,2}-\Delta m_{1}^{2}}%
\right) ^{2}\right] ^{-\frac{1}{2}}
\end{equation}%
Substituting the Eq.(38) into the Eq.(37), we obtain 
\begin{equation}
U_{e1,2}^{m}\approx \frac{\lambda _{1,2}-\Delta m_{1}^{2}}{\left| \lambda
_{1,2}-\Delta m_{1}^{2}\right| }\frac{\alpha _{1,2}\sqrt{\eta _{1}^{2}+\eta
_{2}^{2}}}{\sqrt{\alpha _{1,2}^{2}+\left( \eta _{1}^{2}+\eta _{2}^{2}\right)
\delta _{l}^{2}}}
\end{equation}%
Where 
\begin{equation}
\alpha _{1,2}=\sqrt{\eta _{1}^{2}+\eta _{2}^{2}}\left( \lambda _{1,2}-\frac{%
\eta _{1}^{2}-\eta _{2}^{2}}{\eta _{1}^{2}+\eta _{2}^{2}}\Delta
m_{1}^{2}\right)
\end{equation}%
Using Eq.(34) of $\lambda _{1,2}$, we get 
\begin{equation}
\alpha _{1,2}=\sqrt{\eta _{1}^{2}+\eta _{2}^{2}}\left( \frac{1}{2}\rho
_{l}\mp \frac{1}{2}\sqrt{\rho _{l}^{2}+4\delta _{l}^{2}}\right)
\end{equation}%
Then 
\begin{equation}
U_{e1,2}^{m}\approx \frac{\lambda _{1,2}-\Delta m_{1}^{2}}{\left| \lambda
_{1,2}-\Delta m_{1}^{2}\right| }\frac{\alpha _{1,2}}{\left| \alpha
_{1,2}\right| }\sqrt{\frac{\eta _{1}^{2}+\eta _{2}^{2}}{2}\left( 1\mp \frac{%
\rho _{l}}{\sqrt{\rho _{l}^{2}+4\delta _{l}^{2}}}\right) }
\end{equation}%
The $\eta _{3}^{m}$ is not independent. It is determined by $%
U_{e1}^{m}+U_{e2}^{m}+U_{e3}^{m}=1.$ from it, we obtain 
\begin{equation}
U_{e3}^{m}\approx \eta _{3}
\end{equation}%
When the resonance condition $\rho _{l}=0$ is satisfied, we have 
\begin{equation}
U_{e1}^{m}\approx \frac{\lambda _{1,2}-\Delta m_{1}^{2}}{\left| \lambda
_{1,2}-\Delta m_{1}^{2}\right| }\frac{\alpha _{1,2}}{\left| \alpha
_{1,2}\right| }\sqrt{\frac{\eta _{1}^{2}+\eta _{2}^{2}}{2}}
\end{equation}%
{\bf \vspace{10pt}}

\subsection{Resonant form for $\protect\lambda _{2}$ and $\protect\lambda %
_{3}$ around $A\sim \Delta m_{2}^{2}$.\ \protect\vspace{10pt}}

We have obtained three solutions when $A\sim \Delta m_{2}^{2}$. They are
expressed in Eq.(27) and Eq.(30). There is a possible resonance between
eigenstates of $\lambda _{2}$ and $\lambda _{3}$ when the quantity in the
square root takes a minimum value. It is easy to show that the minimum can
be obtained when the following condition of $A=A_{h}$ is satisfied: 
\begin{equation}
A_{h}=\Delta m_{2}^{2}\left( \eta _{11}+\eta _{22}-\eta _{33}\right)
\end{equation}%
Let 
\begin{equation}
\rho _{h}=A-\Delta m_{2}^{2}\left( \eta _{11}+\eta _{22}-\eta _{33}\right)
\quad and\quad \delta _{h}=\Delta m_{2}^{2}\sqrt{\eta _{11}+\eta _{22}}\eta
_{3}
\end{equation}%
the $\lambda _{2,3}$ can be rewrite as 
\begin{equation}
\lambda _{2,3}=\frac{1}{2}\left( A+\Delta m_{2}^{2}\mp \sqrt{\rho
_{h}^{2}+4\delta _{h}^{2}}\right)
\end{equation}%
It is clear that $4\delta _{h}^{2}>0$ when $\left( \eta _{11}+\eta
_{22}\right) \eta _{33}\neq 0.$ We obtain a possible resonance. Because $%
\lambda _{1}\ll \Delta m_{2}^{2}\sim \lambda _{2,3}$,\ this resonance
between eigenstates of $\lambda _{2}$ and $\lambda _{3}$ is decoupled
approximately from $\lambda _{1}.$

When we take the traditional representation and let $U=U_{2}U_{3}U_{1},$ we
have 
\begin{equation}
\eta _{1}=c_{1}c_{3},\;\eta _{2}=s_{1}c_{3}\;and\;\eta _{3}=s_{3}
\end{equation}%
Then, 
\begin{equation}
\eta _{11}+\eta _{22}-\eta
_{33}=c_{1}^{2}c_{3}^{2}+s_{1}^{2}c_{3}^{2}-s_{3}^{2}=\cos 2\theta _{3}
\end{equation}%
The higher resonance condition, Eq.(45), becomes 
\begin{equation}
A_{h}=\Delta m_{2}^{2}\cos 2\theta _{3}+O\left( \Delta m_{1}^{2}\right)
\end{equation}%
It is the same as the result obtained by Kuo {\it et al.} at large $A$ resonance.%
\vspace{10pt}

\subsection{The behaviors of $U_{e1}^{m}$,\ $U_{e2}^{m}$\ and $U_{e3}^{m}$\
in the neighborhood of $A\sim \Delta m_{2}^{2}$.\protect\vspace{10pt}}

When $A\sim \Delta m_{2}^{2},$ the higher resonance, $\lambda _{1}\ll
\lambda _{2,3}$. We can use the equations of $U_{e2}^{m}$ and $U_{e3}^{m}$
and make the approximation as follows 
\begin{equation}
U_{e2,3}^{m}\approx N^{2,3}\left[ \frac{\eta _{1}^{2}}{\lambda _{2,3}}+\frac{%
\eta _{2}^{2}}{\lambda _{2,3}}+\frac{\eta _{3}^{2}}{\lambda _{2,3}-\Delta
m_{2}^{2}}\right]
\end{equation}%
Correspondingly, we have 
\begin{equation}
N^{2,3}\approx \left[ \left( \frac{\eta _{1}}{\lambda _{2,3}}\right)
^{2}+\left( \frac{\eta _{2}}{\lambda _{2,3}}\right) ^{2}+\left( \frac{\eta
_{3}}{\lambda _{2,3}-\Delta m_{2}^{2}}\right) ^{2}\right] ^{-\frac{1}{2}}
\end{equation}%
Substituting the Eq.(52) into the Eq.(51), we obtain 
\begin{equation}
U_{e2,3}^{m}\approx \frac{\lambda _{2,3}\left( \lambda _{2,3}-\Delta
m_{2}^{2}\right) }{\left| \lambda _{2,3}\left( \lambda _{2,3}-\Delta
m_{2}^{2}\right) \right| }\frac{\alpha _{2,3}}{\sqrt{\alpha
_{2,3}^{2}+\delta _{h}^{2}}}
\end{equation}%
Where 
\begin{equation}
\alpha _{2,3}=\lambda _{2,3}-\Delta m_{2}^{2}\left( \eta _{1}^{2}+\eta
_{2}^{2}\right)
\end{equation}%
Using Eq.(47) of $\lambda _{2,3},$ we get 
\begin{equation}
\alpha _{2,3}=\frac{1}{2}\rho _{h}\mp \frac{1}{2}\sqrt{\rho _{h}^{2}+4\delta
_{h}^{2}}
\end{equation}%
Then 
\begin{equation}
U_{e2,3}^{m}\approx \frac{\lambda _{2,3}\left( \lambda _{2,3}-\Delta
m_{2}^{2}\right) }{\left| \lambda _{2,3}\left( \lambda _{2,3}-\Delta
m_{2}^{2}\right) \right| }\frac{\alpha _{2,3}}{\left| \alpha _{2,3}\right| }%
\sqrt{\frac{1}{2}\left( 1\mp \frac{\rho _{h}}{\sqrt{\rho _{h}^{2}+4\delta
_{h}}}\right) }
\end{equation}%
The $U_{e1}^{m}$ is not independent. It is determined by $%
U_{e1}^{m}+U_{e2,3}^{m}+U_{e3}^{m}=1.$ From it, we obtain 
\begin{equation}
U_{e1}^{m}\approx 0
\end{equation}%
When the resonance condition $\rho _{h}=0$ is satisfied, we have 
\begin{equation}
U_{e2,3}^{m}\approx \frac{\lambda _{2,3}\left( \lambda _{2,3}-\Delta
m_{2}^{2}\right) }{\left| \lambda _{2,3}\left( \lambda _{2,3}-\Delta
m_{2}^{2}\right) \right| }\frac{\alpha _{2,3}}{\left| \alpha _{2,3}\right| }%
\sqrt{\frac{1}{2}}
\end{equation}%
\vspace{10pt}

\section{Discussion and conclusion}

In situation of three neutrinos in matter, the mixing is generally different
to two neutrinos. They do not occur in any plane spanned by two of the three 
eigenvectors $\mid \nu ^{m,u}\rangle $ $(u=1,2,3)$. We can
not use a simple mixing angle to describe their mixing as done in two
neutrino situation. However in case of $\Delta m_{1}^{2}\ll \Delta m_{2}^{2},
$ we have two approximately decoupled resonances. In each case the
oscillation do occurs between two mass eigenvectors, but there is important
difference from two neutrinos situation. 

Around $A\sim \Delta m_{1}^{2}$, matter effect make $U_{e1}^{m}$ and $U_{e2}^{m}$ variation when A change.
The variation has oscillation characteristic and resonance. Oscillation
occurs between $\mid \nu ^{m,1}\rangle $ and $\mid \nu ^{m,2}\rangle $ but
it has an amplitude $\sqrt{U_{e1}^{m}+U_{e2}^{m}}=\sqrt{\eta _{1}^{2}+\eta
_{2}^{2}}<1.$ This amplitude is determined by the vacuum mixing parameters.
In plane spanned by $\mid \nu ^{m,1}\rangle $ and $\mid \nu ^{m,2}\rangle ,$
there exist a mixing angle $\beta _{l}$ which varies with $A$. The $U_{e1}^{m}=a\cos
\beta _{l}$ and $U_{e2}^{m}=a\sin \beta _{l}$ with $a=\sqrt{\eta
_{1}^{2}+\eta _{2}^{2}}<1$. In this meaning, we have oscillation and resonance
in a plane spanned by $\mid \nu ^{m,1}\rangle $ and $\mid \nu ^{m,2}\rangle $%
. It constitutes only a part of $\mid \nu _{e}\rangle .$ The $\mid \nu
^{m,3}\rangle $ contribution is nonzero and has a approximately constant
mixing coefficient $U_{e3}^{m}\mid \nu ^{m,3}\rangle \simeq \eta _{3}\mid
\nu ^{m,3}\rangle .$ Noticeably $\mid \nu ^{m,3}\rangle $ enter only due to
vacuum mixing but not matter effect. It is not oscillation with the 
potential $A$ change. When $U_{e3}=\eta _{3}$ is not small, the nature is
important. It is different with two neutrino oscillation and resonance in
matter.

Around $A\sim \Delta m_{2}^{2}$, the case is same. However, because of the small mass,
the mixing contribution of $\mid \nu ^{m,1}\rangle $ is negligible to $U_{e2}^m$
and $U_{e3}^m$ oscillation. There exist a angle $\beta _h$ which varies with $A$ in a
plane spanned by the eigenvector $\mid \nu ^{m,2}\rangle $ and $\mid \nu
^{m,3}\rangle .$ The $U_{e2}^m=\cos \beta _h$ and $U_{e3}^m=\sin \beta _h$
with a amplitude 1. The oscillation and resonance occur when potential A
change. In this case, $U_{e1}^m\simeq 0$ whatever the values of $\eta _i$ is
taken. Then the oscillation and resonance is more as one of two
neutrinos.

We have used a theoretical framework developed in a previous paper which is
convenient for dealing with the three flavor neutrino oscillations in
matter. In condition of $\Delta m_{1}^{2}\ll \Delta m_{2}^{2},$
approximately decoupled resonance behaviors is studied and discussed in
details. New interesting physical results are obtained. There are important
differences in resonant phenomena between two flavors and three flavors even
in the case of decoupled resonances, in particular when $\eta_3$ is large. 
In traditional mixing angle description, there is mixing order problem. In 
some case, it may mask the physical characteristic and lead to 
misunderstanding.\vspace{10pt}

\acknowledgements

I would like to express my sincere thanks to Prof. A. S. Hirsch, Prof. T. K.
Kuo and the members of High Energy Theory Group, Physics Department, Purdue
University where part of the work was completed, for the kind hospitality
extended to me during my visit from January to April, 2000. Many fruitful\
discussions with Prof. T. K. Kuo are very helpful to this work. I would like
to think Mr. Hai-Jun Pan and Mr. Tao Tu for their help in doing this work.
The author is supported in part by the National Science Foundation, China.


\end{document}